\documentstyle[preprint,aps]{revtex}
\newcommand{\tr}[1]{\mbox{\rm tr} \left\{ #1 \right\}}
\begin{document}
\draft
\preprint{
\parbox{4cm}{
\baselineskip=12pt
TMUP-HEL-9505\\
UT-720\\
October, 1995\\
\hspace*{1cm}
}}
\title{A viable one-family technicolor model}
\author{N.~Kitazawa
 \thanks{e-mail: kitazawa@musashi.phys.metro-u.ac.jp}}
\address{Department of Physics, Tokyo Metropolitan University,\\
         Hachioji-shi, Tokyo 192-03, Japan}
\author{and}
\author{T.~Yanagida
 \thanks{e-mail: yanagida@danjuro.phys.s.u-tokyo.ac.jp}}
\address{Department of Physics, University of Tokyo,\\
         Bunkyo-ku, Tokyo 113, Japan\\
         and\\
         Institute for Theoretical Physics,
         University of California,\\
         Santa Barbara, CA 93106-4030, USA}
\maketitle
\begin{abstract}
We construct a one-family technicolor model
 which is consistent with the precision experiments
 on the electroweak interaction.
The Majorana mass of the right-handed techni-neutrino is introduced
 and the techni-$U(1)_{B-L}$ symmetry is gauged
 to obtain the correct breaking of the electroweak symmetry.
The tree-level kinetic mixing
 between the techni-$U(1)_{B-L}$ and $U(1)_Y$ gauge bosons
 plays an important role for having the consistent value
 of the $S$ parameter.
\end{abstract}
\newpage

The recent experiments on the electroweak interaction
 give strong constraints on the technicolor theory.
Especially,
 the data on the oblique correction,
 which is parameterized by three parameters
 $S$, $T$, and $U$\cite{peskin-takeuchi},
 directly constrain the scenario of
 the dynamical electroweak symmetry breaking
 by the technicolor interaction.
The naive QCD-like one-family technicolor model has been already excluded,
 since it generally gives large values of $S \geq 0.8$
 for $N_{TC} \geq 2$.
These values are about $3.5$-$\sigma$ or more away
 from the value favored by the experiments
 with a reference point $m_t=175$GeV and $m_H=1$TeV\cite{matsumoto}.
The QCD-like one-doublet model
 gives smaller values of $S \geq 0.2$ for $N_{TC} \geq 2$,
 which are, however, about $1$-$\sigma$ or more away
 from the value favored by the experiments.
Therefore,
 many mechanisms to generate the small value of $S$
 have been considered in the technicolor theory.
The walking technicolor dynamics
 itself\cite{appelquist-triantaphyllou,sundrum-hsu},
 the additional $U(1)$ gauge boson
 which mixes with the electroweak gauge bosons\cite{holdom},
 unusual mass spectrum
 of the techni-fermions\cite{gates-terning,appelquist-terning},
 and exotic quantum numbers of the techni-fermions
 in the electroweak gauge interaction\cite{luty-sundrum},
 have been proposed so far to yield a technicolor model
 with the small $S$ parameter.

In this letter
 we consider the technicolor model
 where the right-handed techni-neutrino
 has a Majorana mass\cite{gates-terning}
 which is the remaining degree of freedom
 of the one-family technicolor model.
Both the left-handed and right-handed techni-neutrino
 must belong to the real representation of the technicolor gauge group
 to have the gauge-invariant Majorana mass,
 while keeping the technicolor interaction vector-like.
Since the smallness of the $S$ parameter
 suggests the small technicolor sector
 (small number of the weak doublets),
 we consider the smallest system.
We assign the techni-leptons to the adjoint representations of $SU(2)_{TC}^L$
 (or the fundamental representations of $SO(3)_{TC}^L$),
 and assign the techni-quarks
 to the fundamental representations of $SU(3)_{TC}^Q$.

Since the techni-leptons
 are in the real representations of the strong $SU(2)_{TC}^L$,
 there is no distinction between
 the Dirac condensate $\langle {\bar N}_R N_L \rangle$
 and the Majorana condensate $\langle N_L N_L \rangle$,
 where $N$ denotes the techni-neutrino.
The Majorana condensate of the left-handed techni-neutrino
 is more favorable than the Dirac condensate,
 because of the presence of the Majorana mass of $N_R$
 \footnote{The Dirac condensate of the techni-electron
           is more favorable than the Majorana condensate
           by virtue of the attractive force
           of the electroweak interaction.}.
There must be some interactions which can assist the Dirac condensate.
We gauge the techni-$U(1)_{B-L}$ symmetry, $U(1)_{B-L}^{TF}$,
 and assume that it is spontaneously broken by the dynamics
 which generates the Majorana mass of $N_R$.
$U(1)_{B-L}^{TF}$ is gauge anomaly free,
 since we have the right-handed techni-neutrino.
The exchange diagrams of the $U(1)_{B-L}^{TF}$ gauge boson
 give an attractive force to the Dirac channel,
 but a repulsive force to the Majorana channel.
As we will see later,
 the mixings between the $U(1)_{B-L}^{TF}$ and electroweak gauge bosons
 play a crucial role for producing the small $S$ parameter.

Our technicolor model is based on the gauge group
 $SU(3)_{TC}^Q \times SU(2)_{TC}^L \times U(1)_{B-L}^{TF}$,
 in which the techni-fermions are transformed as,
\begin{center}
\begin{tabular}{cccc}
 & ~~~$SU(3)_{TC}^Q$~~~ & ~~~$SU(2)_{TC}^L$~~~ & ~~$U(1)_{B-L}^{TF}$~~ \\
$\left( \begin{array}{c}  U_L \\ D_L \end{array} \right)$
 & {\bf 3}        & {\bf 1}        & 1/3    \\
$U_R, \quad D_R$
 & {\bf 3}        & {\bf 1}        & 1/3    \\
$\left( \begin{array}{c}  N_L \\ E_L \end{array} \right)$
 & {\bf 1}        & {\bf 3}        & --1    \\
$N_R, \quad E_R$
 & {\bf 1}        & {\bf 3}        & --1    \\
\end{tabular}
\end{center}
Here $U$ and $D$ denote the techni-quarks
 and $E$ denotes the techni-electron.
The techni-fermions $U$, $D$, $N$, and $E$
 belong to the one family representation of the standard-model gauge group
 $SU(3)_C \times SU(2)_L \times U(1)_Y$.
If one assigns the techni-quarks to the triplets of $SU(2)_{TC}^L$,
 the $SU(2)_{TC}^L$ becomes asymptotic non-free.
In this case
 one needs an extra dynamical assumption
 that the theory has non-trivial ultraviolet fixed point.

One can ask whether
 the Dirac condensate $\langle {\bar N}_R N_L \rangle$
 with the Majorana mass of $N_R$ is really possible or not.
We have found that such condensate really occurs
 by solving the Schwinger-Dyson equation
 in ladder and fixed coupling approximation
 (the detailed calculation will be given in ref.\cite{kitazawa-yanagida}).
Since the $SU(2)_{TC}^L$ is working slowly,
 the fixed coupling approximation is reasonable.
The critical value of the gauge coupling constant
 does not change much from the one in the case of vanishing Majorana mass,
 because relatively high energy dynamics is relevant
 to form the condensate in the walking technicolor
 model\cite{holdom-walking,Akiba-Yanagida,YBM,Appelquist-Wijewardhana}.

How strong the $U(1)_{B-L}^{TF}$ must be
 in order to have the Dirac condensate of the techni-neutrino?
Suppose that
 the $N_R$ has the Majorana mass of $M=200 \sim 300$GeV
 (the same order of the techni-fermion mass scale).
We can calculate the contributions to the vacuum energy
 in the one gauge-boson exchange approximation
 when a constant Dirac or Majorana mass is formed.
We can show that when
\begin{equation}
 {{\alpha_{B-L}^{TF}} \over {m_{B-L}^2}}
   = {{0.3} \over {(250 {\rm GeV})^2}}
   = 4.8 \times 10^{-6} \ {\rm GeV}^{-2},
\end{equation}
 the Dirac condensate is favored for $M \le 300$GeV\cite{kitazawa-yanagida}.
We take, in the present analysis,
 $\alpha_{B-L}^{TF} = 0.3$ and $m_{B-L} = 250$GeV,
 where $m_{B-L}$ denotes the mass of the $U(1)_{B-L}^{TF}$ gauge boson $X$
 \footnote{It is natural to assume that
           the masses $m_{B-L}$ and $M$ are the same order,
           since these masses are expected to be generated
           by the same dynamics.}.
Moreover,
 the difference of the condensation scale
 between the techni-electron and the techni-neutrino
 can be roughly estimated from this vacuum energy calculation.
We obtain the value about $60$GeV,
 which is small in comparison with the techni-fermion mass scale
 $\sim 300$GeV.
This result is consistent with the fact that
 the critical gauge coupling is not much affected by the Majorana mass
 in the analysis of the Schwinger-Dyson equation.
We take the ``constituent mass''
 of the techni-neutrino and the techni-electron
 as $m_N=300$GeV and $m_E=400$GeV, respectively,
 in the following numerical calculation.
Notice that
 the $\langle N_L N_L \rangle$ and $\langle N_R N_R \rangle$ condensations
 are disfavored,
 since the $U(1)_{B-L}^{TF}$ interaction acts as a repulsive force
 in these channels.

We should note that
 $U(1)_Y$ and $U(1)_{B-L}^{TF}$ is not ``diagonal'',
 namely, $\tr{Q_Y Q_{B-L}^{TF}} \ne 0$.
Therefore, the bare kinetic mixing term
\begin{equation}
 {\cal L}^{mix} = \omega F_Y{}^{\mu\nu} F_X{}_{\mu\nu}
\label{tree-mixing}
\end{equation}
 must be introduced so that the theory is renormalizable.
Although the new parameter $\omega$
 may be defined in a more fundamental theory,
 we treat it as a free parameter in this letter.
We take $\omega=0.07$ in the following numerical calculations.
This parameter plays an important role for having small $S$ parameter
 \footnote{A similar mixing between
           the additional $U(1)$ and the electroweak gauge bosons
           has been also considered by Holdom\cite{holdom}.}.

Now we turn to discuss
 the compatibility of this model with the precision experiments.

The tree-level mixing in eq.(\ref{tree-mixing})
 yields the tree-level contributions to the $S$, $T$, and $U$ parameters.
By diagonalizing the kinetic and the mass matrices
 of the third component of the $SU(2)_L$, $U(1)_Y$,
 and $U(1)_{B-L}^{TF}$ gauge fields,
 we obtain the following tree-level contributions:
\begin{eqnarray}
 S^{\rm tree}
  &=& {{16} \over \alpha} {{(c^2-r^2) s^2 c^2 \omega^2} \over {(r^2-1)^2}}
   \simeq -0.28,
\label{tree-S}
\\
 T^{\rm tree}
  &=& - {4 \over \alpha} {{r^2 s^2 \omega^2} \over {(r^2-1)^2}}
   \simeq -0.10,
\\
 U^{\rm tree}
  &=& {{16} \over \alpha} {{ s^4 c^2 \omega^2} \over {(r^2-1)^2}}
   \simeq 0.0096,
\end{eqnarray}
 where $r = m_{B-L}/m_Z$,
 and $c$ and $s$ are the cosine and sine of the Weinberg angle,
 respectively.
There are rather large negative contributions
 to the $S$ and $T$ parameters\cite{holdom}.

In addition to the oblique correction,
 the normalization of the neutral current
 and the Weinberg angle are shifted due to the mixing.
The low-energy effective four-fermion interactions
 generated by both the $Z$ and the $U(1)_{B-L}^{TF}$ gauge boson exchanges
 are (following the notation of ref.\cite{kennedy-lynn})
\begin{equation}
 {\cal L}_{eff}^{neuteral} = {1 \over {m_Z^2}} J^f_\mu J^{f'\mu},
\end{equation}
\begin{equation}
 J^f_\mu = {{e_*} \over {c_* s_*}} \sqrt{Z_*}
           {\bar f} \gamma_\mu \left( I_3 - s_*^2 Q \right) f,
\end{equation}
 where $f$ and $f'$ are the ordinary quarks and leptons.
The shifts from the standard model,
 $\delta Z_* = Z_* - Z_*\vert_{SM}$
 and $\delta s_*^2 = s_*^2 - s_*^2\vert_{SM}$,
 are given by\cite{holdom}
\begin{eqnarray}
 \delta Z_*   &=& {{4 r^2 s^2 \omega^2} \over {(r^2-1)^2}}
              \simeq 7.9 \times 10^{-4},
\label{Z-shift}\\
 \delta s_*^2 &=& {{4 s^2 c^2 \omega^2} \over {r^2-1}}
              \simeq 5.3 \times 10^{-4}.
\label{s-shift}
\end{eqnarray}
These shifts are detectable in principle
 by comparing the data at Z-pole, where the Z boson exchange dominates,
 with the data of low energy neutral current experiments,
 $\nu_\mu$-$q$ scattering, $\nu_\mu$-$e$ scattering, and so on.
But these shifts are too small to be detectable
 in the present low energy experiments.

Next we calculate the 1-loop techni-fermion contributions
 to the vacuum polarizations of the electroweak gauge bosons,
 and estimate the contribution to the $S$, $T$, and $U$ parameters.
The mass of the techni-fermion
 is treated as a constant (``constituent mass'').
Since we assume no custodial symmetry breaking in the techni-quark sector,
 the contributions from the techni-quark sector
 to the three parameters are
\begin{eqnarray}
 S^Q &=& {{N_{TC}} \over {6\pi}} \times 3 \simeq 0.48,\\
 T^Q &=& 0,\\
 U^Q &=& 0,
\end{eqnarray}
 where $N_{TC}=3$.
There is a large positive contribution to the $S$ parameter
 as usual in one-family technicolor model.

The contribution from the techni-lepton sector
 is a little complicated,
 because of the Majorana mass of the right-handed techni-neutrino.
The formulae of the techni-lepton contributions
 to the $S$, $T$, and $U$ parameters
 have already been given in ref.\cite{gates-terning}:
\begin{eqnarray}
 S^L &=& {{N_{TC}} \over {6\pi}}
          \Bigg[
           {3 \over 2} + c_M^2 \ln {{m_1^2} \over {m_E^2}}
                       + s_M^2 \ln {{m_2^2} \over {m_E^2}}
\nonumber\\
&& \qquad\qquad\qquad
                       - s_M^2 c_M^2
                        \left({8 \over 3} + f_1(m_1,m_2)
                              -f_2(m_1,m_2) \ln {{m_1^2} \over {m_2^2}}
                        \right)
          \Bigg],\\
 T^L &=& {{N_{TC}} \over {16 \pi s^2 c^2 m_Z^2}}
          \Bigg[
           c_M^2 \left(m_1^2+m_E^2
                       - {{2 m_1^2 m_E^2} \over {m_1^2-m_E^2}}
                         \ln {{m_1^2} \over {m_E^2}}
                 \right)
\nonumber\\
&& \qquad\qquad\quad
         + s_M^2 \left(m_2^2+m_E^2
                       - {{2 m_2^2 m_E^2} \over {m_2^2-m_E^2}}
                         \ln {{m_2^2} \over {m_E^2}}
                 \right)
\nonumber\\
&& \qquad\quad
         	- s_M^2 c_M^2 \left(m_1^2+m_2^2-4 m_1 m_2
                             + 2 {{m_1^3 m_2 - m_1^2 m_2^2 + m_1 m_2^3}
                                   \over
                                  {m_1^2-m_2^2}}
                                  \ln {{m_1^2} \over {m_2^2}}
                       \right)
           \Bigg],\\
 U^L &=& {{N_{TC}} \over {6\pi}}
         \Bigg[
          c_M^2 \left( f_3(m_1,m_E) \ln {{m_1^2} \over {m_E^2}}
                     + {{4 m_1^2 m_E^2} \over {(m_1^2-m_E^2)^2}}
                \right)
\nonumber\\
&& \qquad\qquad\qquad
        + s_M^2 \left( f_3(m_2,m_E) \ln {{m_2^2} \over {m_E^2}}
                     + {{4 m_2^2 m_E^2} \over {(m_2^2-m_E^2)^2}}
                \right)
\nonumber\\
&& \qquad\qquad\qquad
        - {{13} \over 6}
        + s_M^2 c_M^2 \left( {8 \over 3}
                           + f_1(m_1,m_2)
                           - f_2(m_1,m_2) \ln {{m_1^2} \over {m_2^2}}
                      \right)
         \Bigg],
\end{eqnarray}
 where
\begin{eqnarray}
 f_1(m_1,m_2) &=& {{3 m_1 m_2^3 + 3 m_1^3 m_2 - 4 m_1^2 m_2^2}
                    \over
                   {(m_1^2-m_2^2)^2}},\\
 f_2(m_1,m_2) &=& {{m_1^6 - 3 m_1^4 m_2^2 + 6 m_1^3 m_2^3
                          - 3 m_1^2 m_2^4 + m_2^6}
                    \over
                   {(m_1^2-m_2^2)^3}},\\
 f_3(m_1,m_2) &=& {{m_1^6 - 3 m_1^4 m_2^2 - 3 m_1^2 m_2^4 + m_2^6}
                    \over
                   {(m_1^2-m_2^2)^3}},
\end{eqnarray}
\begin{equation}
 m_1 = {{\sqrt{M^2+4 m_N^2} - M} \over 2}, \quad
 m_2 = m_1 + M,
\end{equation}
 $s_M = -\sqrt{m_1/(m_1+m_2)}$, and $c_M = \sqrt{m_2/(m_1+m_2)}$.
Since the Majorana mass breaks the custodial symmetry, 
 we expect a large contribution to the $T$ parameter.
The mass difference
 between the techni-neutrino and the techni-electron due to the Majorana mass
 gives positive contribution to the $T$ parameter,
 but the effect of the Majorana mass itself
 gives the negative contribution to the $T$ parameter.
The negative contribution becomes quite substantial
 when the magnitude of the Majorana mass
 is comparable with the techni-lepton masses.

In total,
 the contribution to the $T$ parameter is
 $0 < T^L < 0.3$ for $M=200 \sim 300$GeV with $m_N=300$GeV and $m_E=400$GeV.
The smaller splitting between $m_N$ and $m_E$
 results in smaller value of $T^L$.
The behavior of the contribution to the $U$ parameter
 is similar to the $T$ parameter,
 but the magnitude is smaller.
Although the Majorana mass
 gives the negative contribution to the $S$ parameter,
 the magnitude is very small,
 when the Majorana mass is comparable with the techni-lepton masses.
The mass splitting between the techni-neutrino and techni-electron
 also gives the negative contribution
 to the $S$ parameter\cite{appelquist-terning},
 but the magnitude is small.
We should stress here that
 the Majorana mass of the right-handed techni-neutrino itself
 does not give an important contribution to have the small $S$ parameter.
Thus,
 our model is completely different from the model
 proposed in ref\cite{gates-terning}.

The mixings between the massive $U(1)_{B-L}^{TF}$
 and the neutral electroweak gauge bosons
 are generated also by the quantum effects.
The exchanges of the $U(1)_{B-L}^{TF}$ gauge boson through the mixings
 (see fig.\ref{diagram-1}) contribute to the vacuum polarizations
 $\Pi_{3Y}'(0)$, $\Pi_{33}(0)$, and $\Pi_{33}'(0)$,
 and change the values of the $S$, $T$, and $U$ parameters,
 where the vacuum polarizations are expanded as
\begin{eqnarray}
 \Pi^{\mu\nu}(q) &=& \Pi(q^2) g^{\mu\nu} + (q^\mu q^\nu \ \mbox{\rm term}),\\
 \Pi(q^2) &=& \Pi(0) + q^2 \Pi'(0) + \cdots.
\end{eqnarray}

The mixings between $U(1)_{B-L}^{TF}$ gauge boson and $W^3$
 which are obtained from the 1-loop diagram of the techni-leptons
 (fig.\ref{diagram-2}) are
\begin{eqnarray}
 \Pi_{3X}(0) &=& - {{N_{TC}} \over {8\pi^2}}
  \Bigg[ 
   c_M^2 (c_M^2-s_M^2) m_1^2 \left(\ln{{\Lambda^2} \over {m_1^2}}-1 \right)
 + s_M^2 (s_M^2-c_M^2) m_2^2 \left(\ln{{\Lambda^2} \over {m_2^2}}-1 \right)
\\
&&
 - s_M^2 c_M^2
   \left\{{{m_1^3 (m_1 - 2 m_2)} \over {m_2^2-m_1^2}}
                    \ln {{\Lambda^2} \over {m_1^2}} 
          - {{m_2^3 (m_2 - 2 m_1)} \over {m_2^2-m_1^2}}
                   \ln {{\Lambda^2} \over {m_2^2}}
          + {1 \over 2} (m_1^2+m_2^2)
   \right\}
  \Bigg],
\nonumber\\
 \Pi_{3X}'(0) &=& {{N_{TC}} \over {16\pi^2}}
                  \Bigg[
                   - {1 \over 3}
                   - {1 \over 3} c_M^2 (c_M^2-s_M^2)
                                 \ln {{m_1^2} \over {m_E^2}}
                   - {1 \over 3} s_M^2 (s_M^2-c_M^2)
                                 \ln {{m_2^2} \over {m_E^2}}
\nonumber\\
&&\qquad
                   + 2 c_M^2 s_M^2
                     \Bigg\{
                      {8 \over 9}
                    + {{m_1 m_2} \over {(m_2^2-m_1^2)^2}}
                       \left( m_1^2 + m_2^2 - {4 \over 3} m_1 m_2 \right)
                    - {2 \over 3} \ln {{m_1 m_2} \over {m_E^2}}
\nonumber\\
&&\qquad
                    - {1 \over 3}
                      {{m_1^6 - 3 m_1^4 m_2^2
                              + 6 m_1^3 m_2^3
                              - 3 m_1^2 m_2^4 + m_2^6}
                        \over
                       {(m_2^2 - m_1^2)^3}}
                      \ln {{m_2^2} \over {m_1^2}}
                     \Bigg\}
                    \Bigg].
\end{eqnarray}
We introduce the ultraviolet cut off
 $\Lambda=1$TeV (scale of the technicolor dynamics)
 in the calculation of the mass mixing $\Pi_{3X}(0)$.
By introducing this physical cut off,
 we approximately include the effect of
 the dumping of the techni-lepton mass function at the scale $\Lambda$.

The mixings
 between the $U(1)_{B-L}^{TF}$ and the $U(1)_Y$ gauge bosons
 are also obtained in the same way as above,
 and given by the followings relations;
\begin{eqnarray}
 \Pi_{YX}(0) &=& - \Pi_{3X}(0),
\label{relation-1}\\
 \Pi_{YX}'(0) &=& - \Pi_{3X}'(0) + 2\omega.
\label{relation-2}
\end{eqnarray}
Note that the kinetic mixing $\Pi_{YX}'(0)$ contains a constant $2\omega$
 which comes from the $\omega$-term in eq.(\ref{tree-mixing}).
The ultraviolet divergence of $\Pi_{YX}'(0)$
 is absorbed by the renormalization of $\omega$.
The simple relations in eqs.(\ref{relation-1}) and (\ref{relation-2})
 are understood by considering the fact that
 $(Y/2)_N = - (I_3)_N$,
 the techni-electron Dirac mass does not break $U(1)_{B-L}^{TF}$ symmetry,
 and $(Y/2)_E = - (I_3)_E + (B-L)_E$.

The correction to the vacuum polarization $\Pi_{33}^{\mu\nu}$
 ($\Pi_{33}(0)$ and $\Pi_{33}'(0)$)
 due to the s-channel $U(1)_{B-L}^{TF}$ gauge boson exchange is given by
\begin{eqnarray}
 \Pi_{33}(0)
   &=& \Pi_{3X}(0) {{~~4\pi\alpha_{B-L}~~} \over {-m_{B-L}^2}} \Pi_{3X}(0),
\\
 \Pi_{33}'(0)
   &=& \Pi_{3X}'(0) {{~~4\pi\alpha_{B-L}~~}
                     \over {-m_{B-L}^2}} \Pi_{3X}(0)
\nonumber\\
   &+& \Pi_{3X}(0) {{~~4\pi\alpha_{B-L}~~}
                     \over {-m_{B-L}^2}} \Pi_{3X}'(0)
\nonumber\\
   &+&  \Pi_{3X}(0) {{~~4\pi\alpha_{B-L}~~}
                                \over {-m_{B-L}^4}} \Pi_{3X}(0).
\end{eqnarray}
These give the positive contributions to the $T$ and $U$ parameters:
\begin{eqnarray}
 T^{B-L} &=& {{4\pi} \over {s^2 c^2 m_Z^2}}
              \left[ \Pi_{11}(0) - \Pi_{33}(0) \right],
\\
 U^{B-L} &=& 16\pi \left[ \Pi_{11}'(0) - \Pi_{33}'(0) \right],
\end{eqnarray}
 where $\Pi_{11}(0)$ and $\Pi_{11}'(0)$ are zero in the present approximation.
The $T$ parameter increases quickly as the Majorana mass becomes larger.
(This is the 2-loop level contribution,
 since $\Pi_{3X}(0)$ is estimated at the 1-loop level.)
When $m_N = 300$GeV and $m_E = 400$GeV, $T^{B-L}<0.35$ for $M<300$GeV.
The behavior of the contribution to the $U$ parameter
 is similar to the $T$ parameter,
 but the magnitude is smaller.

The correction to the vacuum polarization $\Pi_{3Y}'(0)$
 due to the s-channel $U(1)_{B-L}^{TF}$ gauge boson exchange is given by
\begin{eqnarray}
 \Pi_{3Y}'(0) &=& \Pi_{3X}'(0) {{~~4\pi\alpha_{B-L}~~}
                                \over {-m_{B-L}^2}} \Pi_{YX}(0)
\nonumber\\
              &+&  \Pi_{3X}(0) {{~~4\pi\alpha_{B-L}~~}
                                \over {-m_{B-L}^2}} \Pi_{YX}'(0)
\nonumber\\
              &+&  \Pi_{3X}(0) {{~~4\pi\alpha_{B-L}~~}
                                \over {-m_{B-L}^4}} \Pi_{YX}(0).
\end{eqnarray}
Note that the second term contains the 1-loop contribution,
 since $\Pi_{YX}'(0)$ contains the tree-level constant term.
Therefore,
 we have a large contribution to the $S$ parameter:
\begin{equation}
 S^{B-L} = -16 \pi \Pi_{3Y}'(0).
\end{equation}
This contribution is negative taking $\omega$ positive,
 and the magnitude is large enough
 to cancel the large positive contribution
 from the techni-quark sector,
 together with the tree-level contribution in eq.(\ref{tree-S}).
We should stress here that
 this large negative contribution disappears
 when the Majorana mass vanishes,
 since $\Pi_{3X}(0)$ vanishes if $M=0$.
Therefore,
 both the Majorana mass and the $\omega$-term in eq.(\ref{tree-mixing})
 are needed in order to have the large negative contribution.
Holdom has already found that
 the $\omega$-term gives rather large negative contribution
 to the $S$ parameter at tree level.
But the tree-level contribution is not large enough
 to cancel out the large positive value
 in the QCD-like one-family technicolor model,
 while keeping the shifts of eqs.(\ref{Z-shift}) and (\ref{s-shift})
 small\cite{holdom}.

The Majorana mass dependences
 of the total values of the $S$, $T$, and $U$ parameters
 are shown in fig.\ref{S-paramter}, fig.\ref{T-paramter},
 and fig.\ref{U-paramter}, respectively.
All three parameters
 are consistent with the experimental constraints,
 when the Majorana mass of the right-handed techni-neutrino $M<300$GeV.
Remember that we take the parameters $\alpha_{B-L}$ and $m_{B-L}$
 so that the correct electroweak symmetry breaking really occurs
 with $M<300$GeV.
And the mass splitting
 between the techni-neutrino and techni-electron ($100$GeV)
 is a natural one which comes from the estimation of the vacuum energy.
We set the value of $\omega$ to $0.07$
 so that all the things become consistent.
Although
 the value of the $S$ parameter
 may be enhanced by the factor two or more due to non-perturbative effects,
 this model will be still consistent
 by virtue of the large cancelation of the $S$ parameter
 in the region $M \simeq 200$GeV.

We should note that the $T$ parameter is very sensitive
 to the mass deference
 between the techni-neutrino and techni-electron.
If we take smaller mass difference,
 the $T$ parameter becomes negative in the region $M<300$GeV.
If we take the values $m_N=340$GeV and $m_E=400$GeV, for instance,
 the minimum value of $T$ is about $-0.2$ at $M \simeq 250$GeV,
 while the $S$ and $U$ parameters
 are still consistent with the experimental constraints.
Therefore,
 we may explain the deviation of $R_b=\Gamma_b/\Gamma_{had}$
 from the standard-model value
 by considering the effect of the diagonal extended technicolor (ETC)
 gauge boson\cite{wu,hagiwara-kitazawa},
 since the large positive contribution to the $T$ parameter\cite{yoshikawa}
 due to the diagonal ETC boson can be cancelled out.

The number of pseudo-Nambu-Goldstone bosons
 is reduced in comparison with the naive one-family technicolor theory,
 since the approximate chiral symmetry is largely reduced
 by the separate structure of the technicolor gauge group.
If the standard-model gauge interaction is switched off,
 the non-anomalous chiral symmetry of the techni-fermion sector is
 $SU(6)_L^Q \times SU(6)_R^Q \times SU(2)^L_L \times U(1)_V^Q \times U(1)_Y^L$.
Techni-fermion condensates break this chiral symmetry to
 $SU(6)_V^Q \times U(1)_V^Q \times U(1)_{em}^L$,
 and the currents corresponding to the broken symmetries are
\begin{eqnarray}
 J^{ai}_\mu
  &=& {\bar Q} \gamma_\mu \gamma_5 {{\lambda^a} \over 2}{{\tau^i} \over 2} Q
   = F_Q \partial_\mu \Theta^{ai} + \cdots,
\\
 J^a_\mu
  &=& {\bar Q} \gamma_\mu \gamma_5 {{\lambda^a} \over 2} Q
   = F_Q \partial_\mu \Theta^a + \cdots,
\\
 J^{Qi}_\mu
  &=& {\bar Q} \gamma_\mu \gamma_5 {{\tau^i} \over 2} Q
   = F_Q \partial_\mu \Phi_Q^i + \cdots,
\\
 J^{Li}_\mu
  &=& {\bar L} \gamma_\mu {{1-\gamma_5} \over 2} {{\tau^i} \over 2} L
   = F_L \partial_\mu \Phi_L^i + \cdots,
\end{eqnarray}
 where $Q=(U \ D)^T$ and $L=(N \ E)^T$,
 and the last equalities in each equations
 denote the effective couplings of the Nambu-Goldstone bosons
 with decay constants $F_Q$ and $F_L$.
The scales of $F_Q$ and $F_L$ are determined
 by the dynamics of $SU(3)_{TC}^Q$ and $SU(2)_{TC}^L$, respectively.
The true Nambu-Goldstone bosons
 which couple with the electroweak currents are
\begin{equation}
 \Pi^i = \Phi_L^i \cos \varphi - \Phi_Q^i \sin \varphi,
\end{equation}
 and the pseudo-Nambu-Goldstone bosons are $\Theta^{ai}$, $\Theta^a$, and
\begin{equation}
 P^i = \Phi_L^i \sin \varphi + \Phi_Q^i \cos \varphi,
\end{equation}
 where $\tan \varphi = F_Q / F_L$,
 and the decay constant of the true Nambu-Goldstone bosons is
 $\sqrt{F_Q^2+F_L^2}$.
The masses of the colored pseudo-Nambu-Goldstone bosons,
 $\Theta^{ai}$ and $\Theta^a$, are expected to be about $300$GeV.
The electroweak interaction gives the mass to $P^i$.
Although the naive estimation for the mass of $P^i$ is about $10$GeV,
 it will be lifted up by the walking technicolor dynamics of $SU(2)_{TC}^L$.

Finally, we comment on a possible physics beyond the present model.
Because of
 the strong coupling of $U(1)_{B-L}^{TF}$ gauge interaction
 ($\alpha_{B-L}=0.3$ at the electroweak symmetry breaking scale)
 and the presence of many $U(1)_{B-L}^{TF}$ charged techni-fermions,
 the gauge coupling constant for $U(1)_{B-L}^{TF}$ blows up at about $3$TeV.
Therefore,
 we must invoke some new physics in the TeV region.
The technicolor structure may be changed there
 like in the extended technicolor theory
 embedding the $U(1)_{B-L}^{TF}$ in some non-Abelian gauge group.

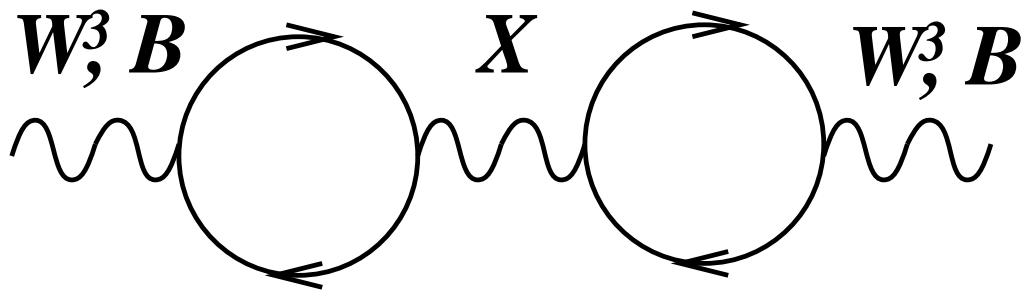
\begin{figure}
\caption{The diagrams which contribute to the vacuum polarizations
         $\Pi_{3Y}'(0)$, $\Pi_{33}(0)$, and $\Pi_{33}'(0)$.
         We consider only the diagrams
         in which the $U(1)_{B-L}^{TF}$ gauge boson $X$
         is exchanged in s-channel.}
\label{diagram-1}
\end{figure}
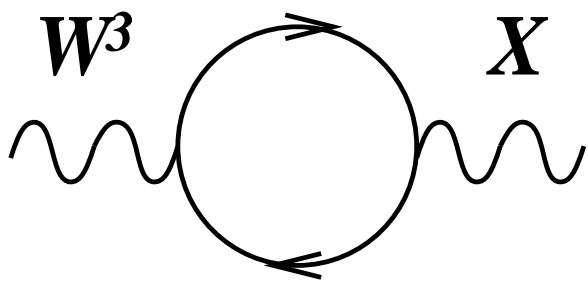
\begin{figure}
\caption{The diagram which gives the mixing
         between the $U(1)_{B-L}^{TF}$ gauge boson $X$ and $W^3$.
         Only the techni-leptons contribute to the loops,
         since only the Majorana mass of the right-handed techni-neutrino
         breaks $U(1)_{B-L}^{TF}$ gauge symmetry.}
\label{diagram-2}
\end{figure}
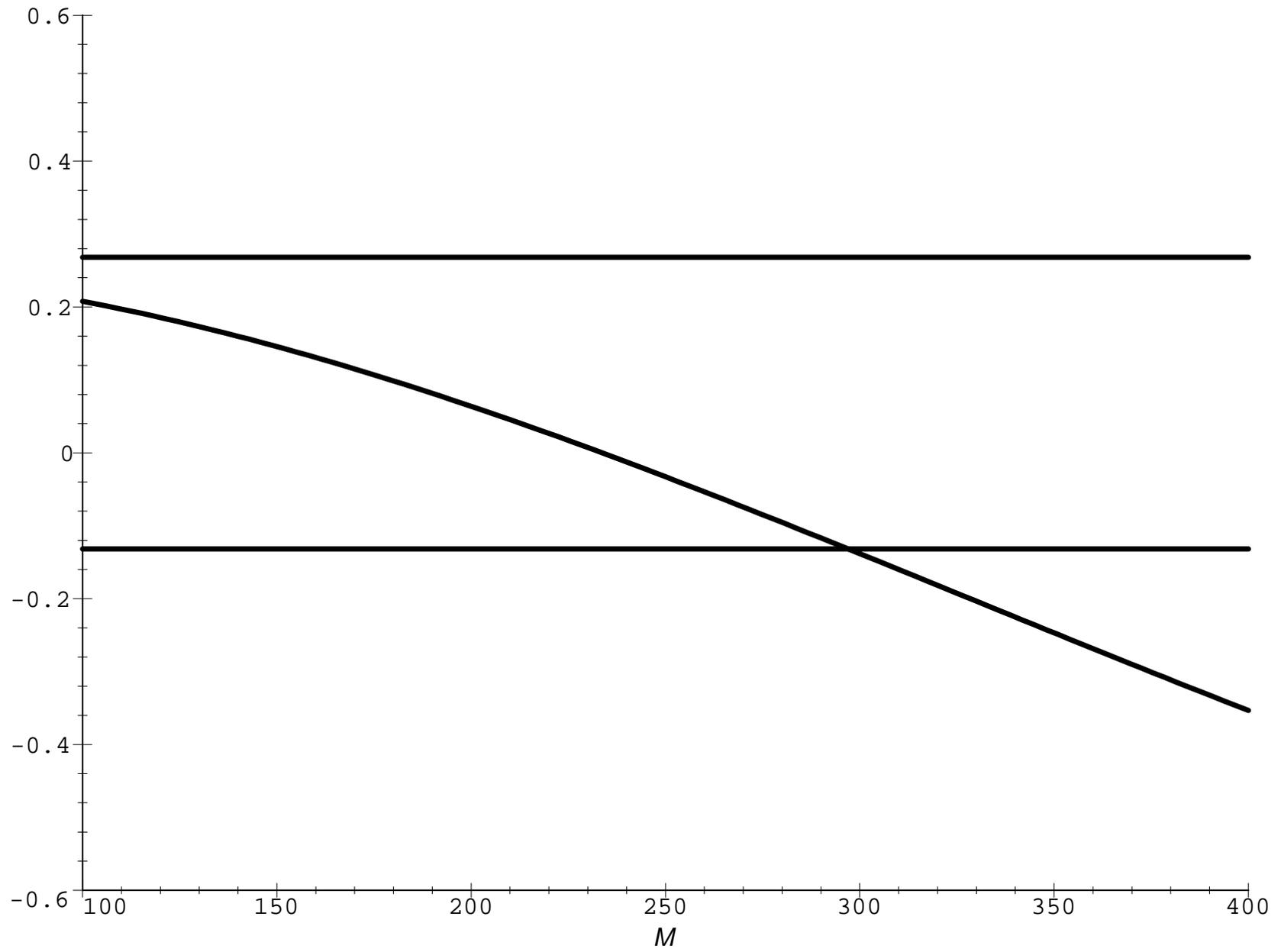
\begin{figure}
\caption{The Majorana mass dependence of the $S$ parameter.
         The region between the two horizontal lines
          is allowed by the experiments [2].
         The reference point is taken as $m_t=175$GeV and $m_H=1$TeV.}
\label{S-paramter}
\end{figure}
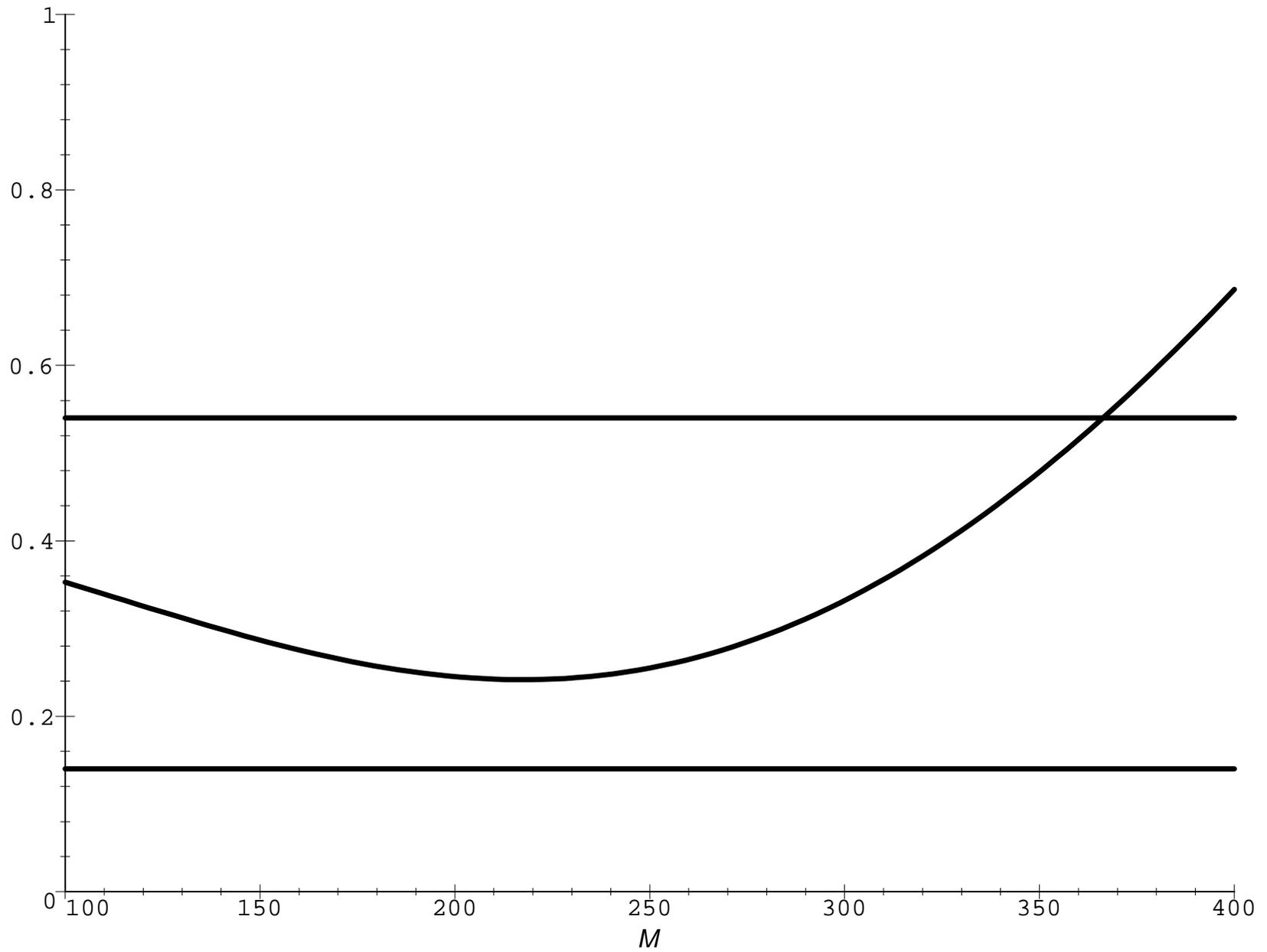
\begin{figure}
\caption{The Majorana mass dependence of the $T$ parameter.
         The region between the two horizontal lines
          is allowed by the experiments [2].
         The reference point is taken as $m_t=175$GeV and $m_H=1$TeV.}
\label{T-paramter}
\end{figure}
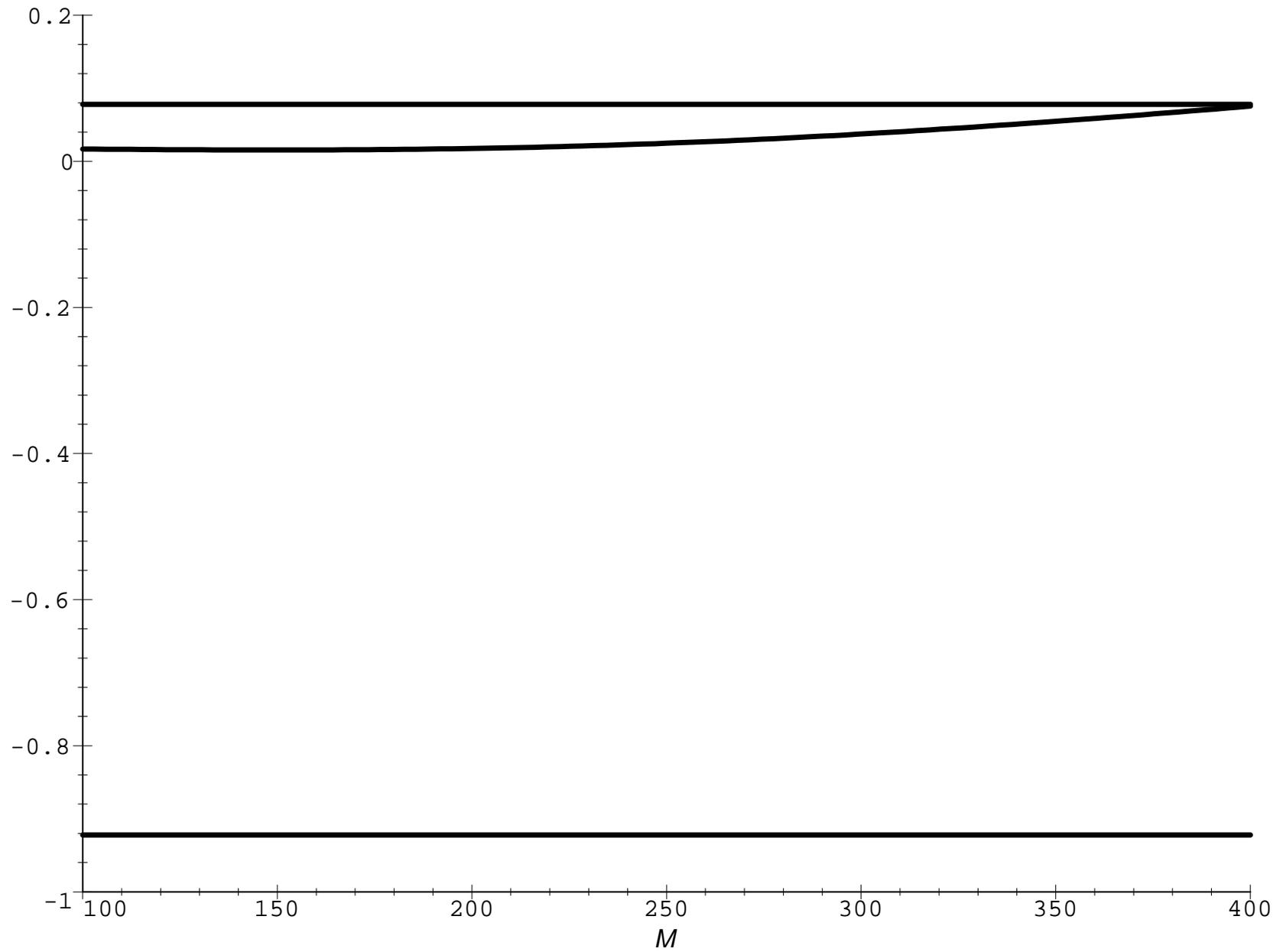
\begin{figure}
\caption{The Majorana mass dependence of the $U$ parameter.
         The region between the two horizontal lines
          is allowed by the experiments [2].
          The reference point is taken as $m_t=175$GeV and $m_H=1$TeV.}
\label{U-paramter}
\end{figure}
\end{document}